\documentclass[aps,twocolumn,superscriptaddress,prb]{revtex4-2}

\usepackage{graphicx}
\usepackage{latexsym}
\usepackage{amssymb}
\usepackage{amsmath}
\usepackage{amsfonts}
\usepackage{upgreek}
\usepackage{subfigure}
\usepackage{bm}
\usepackage{bbold}
\usepackage{verbatim}
\usepackage{hyperref}
\hypersetup{unicode=true, breaklinks=false,pdfborder={0 0 1},colorlinks=true,linkcolor=[rgb]{0,0,1},citecolor=[rgb]{0,0,1},urlcolor=[rgb]{0,0,1}}
\usepackage{multirow}
\usepackage{color}
\usepackage{tikz}
\usepackage{physics}
\usepackage{enumitem}

\newcommand{\f}[1]{\hat{c}_{#1}}
\newcommand{\fd}[1]{\hat{c}_{#1}^\dagger}

\newcommand{\NN}[1]{\left\langle{#1}\right\rangle}

\newcommand{\Si}[1]{\hat{\mathbf{S}}_{#1}}
\newcommand{\Ham}{\hat{\mathcal{H}}}
\newcommand{\iu}{\mathrm{i}\mkern1mu}
\newcommand{\hc}{\mathrm{h.c.}}

\newcommand{\e}[1]{\mathrm{e}^{#1}}
\newcommand{\mb}[1]{\mathbf{#1}}
\newcommand{\mr}[1]{\mathrm{#1}}

\newcommand{\gutz}{\hat{\mathcal{P_G}}}
\newcommand{\tmf}{T_\mathrm{MF}/J_\mathrm{MF}}
\newcommand{\hmf}{h/J_\mathrm{MF}}

\newcommand{\daq}{\delta\xi_l}

\newcommand{\Tbar}{\overline{\Theta}_l}
\newcommand{\Pbar}{\overline{\Phi}_l}
\newcommand{\pare}[1]{\left(#1\right)}
\newcommand{\sqare}[1]{\left[#1\right]}

\newcommand{\dmdaq}{\hat{\rho}_{\xi+\delta\xi_l}}
\newcommand{\dma}{\hat{\rho}_{\xi}}

\begin{document}

\title{Optimized Gutzwiller Projected States for Doped Antiferromagnets in Fermi-Hubbard Simulators}

\author{Christian Reinmoser}
\affiliation{Department of Physics and Arnold Sommerfeld Center for Theoretical Physics (ASC), Ludwig-Maximilians-University Munich, Theresienstr. 37, Munich D-80333, Germany}
\affiliation{Munich Center for Quantum Science and Technology, Schellingstr. 4, Munich D-80799, Germany}
\affiliation{University of Vienna, Faculty of Physics, Boltzmanngasse 5, 1090, Vienna, Austria}

\author{Muqing Xu}
\affiliation{
Department of Physics, Harvard University, 17 Oxford St., Cambridge, Massachusetts 02138, USA
}

\author{Lev Haldar Kendrick}
\affiliation{
Department of Physics, Harvard University, 17 Oxford St., Cambridge, Massachusetts 02138, USA
}

\author{Anant Kale}
\affiliation{
Department of Physics, Harvard University, 17 Oxford St., Cambridge, Massachusetts 02138, USA
}

\author{Youqi Gang}
\affiliation{
Department of Physics, Harvard University, 17 Oxford St., Cambridge, Massachusetts 02138, USA
}

\author{Martin Lebrat}
\affiliation{
Department of Physics, Harvard University, 17 Oxford St., Cambridge, Massachusetts 02138, USA
}
\affiliation{JILA, National Institute of Standards and Technology and the University of Colorado, Boulder, Colorado 80309-0440, USA}

\author{Markus Greiner}
\affiliation{
Department of Physics, Harvard University, 17 Oxford St., Cambridge, Massachusetts 02138, USA
}

\author{Fabian Grusdt}
\affiliation{Department of Physics and Arnold Sommerfeld Center for Theoretical Physics (ASC), Ludwig-Maximilians-University Munich, Theresienstr. 37, Munich D-80333, Germany}
\affiliation{Munich Center for Quantum Science and Technology, Schellingstr. 4, Munich D-80799, Germany}

\author{Annabelle Bohrdt}
\affiliation{Department of Physics and Arnold Sommerfeld Center for Theoretical Physics (ASC), Ludwig-Maximilians-University Munich, Theresienstr. 37, Munich D-80333, Germany}
\affiliation{Munich Center for Quantum Science and Technology, Schellingstr. 4, Munich D-80799, Germany}
\affiliation{Institute of Theoretical Physics, University of Regensburg, Universitätsstr. 31, Regensburg D-93053, Germany}

\date{\today}

\begin{abstract}
In quantum many-body physics, one aims to understand emergent phenomena and effects of strong interactions, ideally by developing a simple theoretical picture.
Recently, progress in quantum simulators has enabled the measurement of site resolved snapshots of Fermi-Hubbard systems at finite doping on square as well as triangular lattice geometries. 
These experimental advances pose the quest for theorists to analyze the ensuing data in order to gain insights into these prototypical, strongly correlated many-body systems.
Here we employ machine learning techniques to optimize the mean-field parameters of a resonating valence bond (RVB) state through comparison with experimental data, thus determining a possible underlying simple model that is physically motivated and fully interpretable.
We find that the resulting RVB states are capable of capturing two- as well as three-point correlations measured in experiments, even when they are not specifically used in the optimization.
The analysis of the mean-field parameters and their doping dependence can be used to obtain physical insights and shed light on the nature of possible underlying quantum spin liquid states. 
Our results show that finite temperature data from Fermi-Hubbard quantum simulators can be well captured by RVB states.
This work paves the way for a new, systematic analysis of data from numerical as well as quantum simulation of strongly correlated quantum many-body systems.
\end{abstract}

\maketitle

\section{Introduction}\label{sec:intro}

Understanding the physics of the doped Fermi-Hubbard model has been an outstanding challenge in condensed matter physics for decades, yielding a variety of theoretical proposals on how to capture the interplay of spin and charge degrees of freedom.
With recent experimental advances in preparing as well as imaging cold fermionic atoms in optical lattices with single site resolution, the need arises to interpret the resulting data in order to gain theoretical insights. 

Shortly after the discovery of high temperature superconductivity in the cuprate materials \cite{Bednorz1986}, Anderson put forward resonating valence bond (RVB) states as a possible description \cite{Anderson1987}, the idea being that such states can potentially capture the physics of a Hubbard model at finite doping, where the hole motion introduces frustration in the spin sector. 

Anderson's RVB picture \cite{Anderson1987} considers trial wavefunctions of free 'holons' moving through a spin liquid comprised of singlet coverings. Originally, Anderson suggested RVB states as a possible ground state of the two dimensional Heisenberg model on a triangular lattice \cite{Anderson1973}. It can be shown that if the total state of a system is a spin singlet, the state can be written as a superposition of states, each of which corresponds to some pairwise singlet configuration covering. Resonating valence bond states describe states which are superpositions of different coverings of the lattice with singlets, typically on nearest neighbor bonds. 

An RVB state as trial wavefunction or variational ansatz can be represented as $\ket{\psi} = \gutz \ket{\psi_0}$ \cite{Gutzwiller1963}, where $\ket{\psi_0}$ is the ground state of a fermionic mean field Hamiltonian \cite{Gros1989}, which qualitatively captures low-energy features of interest. The state $\ket{\psi}$ thus fulfills directly the fermionic antisymmetry properties and can at the same time be evaluated efficiently both at half-filling and at finite doping. There are only few variational parameters in this ansatz, namely the mean field parameters. The Gutzwiller projection $ \gutz$ projects out double occupancies, physically motivated by the limit of large interaction strength $U\gg t$ in the Hubbard Hamiltonian, and thus builds in correlations in the state $\ket{\psi}$.   

Early studies \cite{Paramekanti2001,Paramekanti2004} have shown that a Gutzwiller projected $d$-wave state can qualitatively capture various experimental observations in the cuprate materials \cite{Anderson2004}, such as the doping dependence of the nodal Fermi velocity and the quasiparticle weight, as well as the suppression of the Drude weight and the superfluid density \cite{Edegger2007}. Other low-energy aspects, like the Fermi arcs observed in the pseudogap phase or the competition of the stripe phase with superconductivity \cite{Qin2020,Xu2024}, cannot be captured by the RVB ansatz. Nevertheless, it remains a powerful approach for capturing the general behavior of correlation functions at elevated temperatures \cite{Chiu2019,Bohrdt2019,Koepsell2021}.

We aim to capture the local features of the doped Fermi-Hubbard model at finite temperatures, measured using ultracold atoms, on the square as well as the triangular lattice with RVB states. We optimize the mean field parameters describing our finite temperature RVB states by gradient descent through comparison to experimental data from quantum gas microscopes \cite{Gross2017,Bohrdt2021a}. This allows an interpretation of the data in terms of a doped spin liquid while identifying which types of spin liquids may be incompatible with experiments.

\begin{figure}[htbp]
    \centering
    \includegraphics[width=\linewidth]{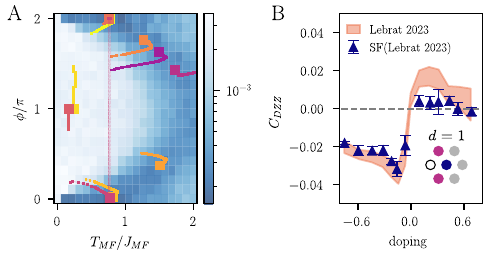}
    \caption{\textbf{Optimizing finite temperature RVB states through comparison to experimental data.} 
Panel \textbf{A} depicts the objective function with respect to data from the square lattice realization of the Fermi-Hubbard model in Ref.~\cite{Chiu2019} at half-filling as a function of parameters $\phi$ and $\tmf$ with $\hmf=0$. We observe an arc-shaped region with low values for the objective function $\mathcal{L}$, see Eq.~\eqref{eq:ObjFun}, in which we usually end up after our optimization routine (final parameters marked by squares). Yet, there are two runs which did not converge properly in the white region.
The line spanning the entire $\phi$-axis corresponds to an optimization run where $\phi$ became negative which is then mapped to $2\pi-\abs{\phi}$.
 \textbf{B} shows the spin-spin correlation function next to a dopant $C_\mr{DZZ}$ for the triangular lattice Fermi-Hubbard model averaged over ten optimization runs with varying initial conditions. We compare $C_\mr{DZZ}$ obtained from the optimized RVB to the experimental data from \cite{Lebrat2024}. We observe that the SF states are capable of capturing the correlator adequately in the hole doped regime, while underestimating it in the particle doped regime. Nevertheless, the sign change for the connected correlator \(C_{DZZ}\) can be reproduced.
 }%
    \label{fig:intro}
  \end{figure}

Quantum simulators are starting to reach regimes which are challenging to simulate numerically and to understand theoretically, for instance the two-dimensional (2D) Hubbard model at finite doping and finite temperature, as well as potential 2D quantum spin liquid states \cite{Semeghini2021}. In this paper, we demonstrate a systematic procedure to obtain the best possible description in terms of resonating valence bond states based on a Gutzwiller projected finite temperature mean field description, see Fig.~\ref{fig:intro}.  

The remainder of this paper is organized as follows: in Sec.~\ref{sec:model}, we introduce the considered models as well as our RVB ansatz. In Sec.~\ref{sec:opti}, we discuss our gradient descent based optimization method, which we benchmark in Sec.~\ref{sec:bench}. We then analyze the experimental data for the square lattice in Sec.~\ref{sec:square} and for the triangular lattice in Sec.~\ref{sec:tri}, and conclude in Sec.~\ref{sec:sum}.

\section{Model and Trial States}\label{sec:model}

We consider the Fermi-Hubbard model,
\begin{align}
\begin{split}
    \hat{\mathcal{H}}_\mathrm{FH} = \
    &- t \sum_{\langle \mathbf{i}, \mathbf{j} \rangle, \sigma} (\hat{c}^\dagger_{\mathbf{i}, \sigma} \hat{c}_{\mathbf{j}, \sigma} + \hc)
    + U \sum_\mathbf{i} \hat{n}_{\mathbf{i} \uparrow} \hat{n}_{\mathbf{i} \downarrow}
\end{split}
\label{eq:Fermi-Hubbard}
\end{align}
with tunneling $t$ and on-site interaction $U$ in a spin-balanced setting. Here, $\hat{c}^{(\dagger)}_{\mathbf{i}, \sigma}$ are the fermionic annihilation (creation) operators at site $\mathbf{i} = (x, y)$ with spin $\sigma$ and $\hat{n}_\mathbf{i} = \sum_\sigma \hat{n}_{\mathbf{i}, \sigma} = \sum_\sigma \hat{c}^{\dagger}_{\mathbf{i}, \sigma} \hat{c}_{\mathbf{i}, \sigma}$ is the particle number operator.
In the limit of strong interactions $U \gg t$, the Fermi-Hubbard model can -- up to three-site terms -- in first order in $t/U$ be approximated by the $t-J$ model
  \begin{align}\label{eq:tJHam}
   \begin{split}
      \Ham_{\mathrm{tJ}} = & -t~\gutz \left( \sum_{\NN{\mb{i},\mb{j}},\sigma} \hat{c}^\dagger_{\mathbf{i}, \sigma} \hat{c}_{\mathbf{j}, \sigma} + \hc \right) \gutz \\
      & + J \sum_{\NN{\mb{i},\mb{j}}} \left( \Si{\mb{i}} \cdot \Si{\mb{j}} - \frac{\hat{n}_\mb{i}\hat{n}_\mb{j}}{4} \right),
   \end{split}
  \end{align}
  where $\gutz$ projects onto singly occupied sites and $\Si{i}$ are the spin operators. 
In order to obtain a RVB ansatz, we derive a mean field Hamiltonian \cite{Wen2002}, where we express the spin operators as
\begin{equation}
\hat{\mathbf{S}}_\mathbf{i} = \frac{1}{2} \sum_{\alpha, \beta}  \hat{c}_{\mathbf{i},\alpha}^\dagger \boldsymbol{\sigma}_{\alpha,\beta} \hat{c}_{\mathbf{i},\beta}
\end{equation}
with $\boldsymbol{\sigma}$ the vector of Pauli matrices. 
We then perform a mean field decoupling and replace the operators $\hat{c}_{\mathbf{i},\alpha}^\dagger \hat{c}_{\mathbf{j},\alpha}$ and $\hat{c}_{\mathbf{i},\alpha}^\dagger \hat{c}_{\mathbf{i},\alpha}$ by their expectation values, yielding the mean field Hamiltonian
  \begin{align}\label{eq:MFHam}
   \begin{split}
      \Ham_\mr{MF}=&-J_\mr{MF}\sum_{\NN{\mb{i},\mb{j}},\sigma}\left( \e{\iu \phi_{\mb{i}\mb{j}}} \fd{\mb{i},\sigma} \f{\mb{j},\sigma} + \hc \right)\\
      & - h \sum_{\mb{i},\sigma} \sigma \e{\iu \mathbf{Q}\cdot\mb{R}_\mb{i}} \fd{\mb{i},\sigma} \f{\mb{i},\sigma},
   \end{split}
  \end{align}

where $\mb{Q}=(\pi,\pi)$ and $\mb{R}_\mb{i}$ is the position of the particle and the first sum is over nearest neighbor pairs.
The ansatz state we will use in the remainder of the paper is then a Gutzwiller projected finite temperature state of this mean field Hamiltonian \eqref{eq:MFHam}.

We analyze the model on the square and the triangular lattice.
For the square lattice the nearest neighbors are defined as \(\mb{j} = \mb{i} + \mb{x}\) and \(\mb{j} = \mb{i} + \mb{y}\), where \(\mb{x},\mb{y}\) are the primitive translation vectors of the lattice. In this case we choose the following parameters for the staggered flux state:
\begin{equation*}
    \phi_{\mb{i}\mb{j}} =
    \begin{cases}
        +\frac{\phi}{4}& \text{for} \ \norm{\mb{i}}_1~\text{odd},~\mb{j}=\mb{i}+\mb{x} \\
        +\frac{\phi}{4}& \text{for} \ \norm{\mb{i}}_1~\text{even},~\mb{j}=\mb{i}+\mb{y} \\
        -\frac{\phi}{4}&\text{for}~\norm{\mb{i}}_1~\text{even},~\mb{j}=\mb{i}+\mb{x} \\
        -\frac{\phi}{4}&\text{for}~\norm{\mb{i}}_1~\text{odd},~\mb{j}=\mb{i}+\mb{y}.
    \end{cases}
\end{equation*}
where \(\norm{\cdot}_1\) is the \(1\)-norm.
We analyze the triangular lattice in a similar fashion by considering additional diagonal interaction in a square lattice, such that the nearest neighbors include \(\mb{j} = \mb{i} + \mb{x}\), \(\mb{j} = \mb{i} + \mb{y}\) and \(\mb{j} = \mb{i} + \mb{x} + \mb{y}\).
The parameters for the staggered flux state are then
\begin{equation*}
    \phi_{\mb{i}\mb{j}} =
    \begin{cases}
        +\frac{\phi}{4}& \text{for} \ \norm{\mb{i}}_1~\text{odd},~\mb{j}=\mb{i}+\mb{x} \\
        +\frac{\phi}{4}& \text{for} \ \norm{\mb{i}}_1~\text{even},~\mb{j}=\mb{i}+\mb{y} \\
        -\frac{\phi}{4}&\text{for}~\norm{\mb{i}}_1~\text{even},~\mb{j}=\mb{i}+\mb{x} \\
        -\frac{\phi}{4}&\text{for}~\norm{\mb{i}}_1~\text{odd},~\mb{j}=\mb{i}+\mb{y},\\
        -\frac{\phi}{2}&\text{for}~\norm{\mb{i}}_1~\text{even},~\mb{j}=\mb{i}+\mb{x}+\mb{y}\\
        +\frac{\phi}{2}&\text{for}~\norm{\mb{i}}_1~\text{odd},~\mb{j}=\mb{i}+\mb{x}+\mb{y}.
    \end{cases}  
\end{equation*}
We note that for the analysis of the experimental data and all calculations regarding the triangular lattice, we choose \(h=0\).

Diagonalizing the quadratic Hamiltonian $\Ham_\mr{MF}$ yields Slater-determinant eigenstates $\ket{\alpha_\mb{k}}$ with eigenenergies $E_{\alpha_\mb{k}}$. A finite temperature mean field state $\hat{\rho}_\mr{MF}$ at temperature $\beta_\mr{MF}$ can thus be directly constructed by occupying the states $\ket{\alpha_\mb{k}}$ according to the thermal distribution. 
We apply an extension of well-established methods from variational Monte Carlo \cite{Becca2017} to obtain estimates for expectation values of Gutzwiller projected thermal states \cite{Chiu2019} with total spin $S=0$. In particular, we directly apply the Gutzwiller projection $\gutz$ by only allowing singly occupied sites in the real space configurations $\ket{\alpha_\mb{r}}$ during the Monte Carlo sampling procedure, thus accessing the state

  \begin{equation}\label{eq:finTDM}
    \hat{\rho} = \frac{1}{Z}~\gutz \left( \sum_{\alpha_\mb{k}} \ket{\alpha_\mb{k}} \bra{\alpha_\mb{k}} \e{-\beta_\mr{MF} E_{\alpha_\mb{k}}} \right) \gutz,
  \end{equation}
where $Z$ acts as a normalization constant.
Our trial state, given in Eq. \eqref{eq:finTDM}, depends on the dimensionless parameters $T_\mr{MF}/J_\mr{MF}$, $\phi$ and $h/{J_\mr{MF}}$ and we refer to it as staggered flux and Néel (SFN) state or staggered flux (SF) state if $\hmf=0$.
The temperature dependence of the state is entirely contained in the exponent as $\beta_\mr{MF} = {1}/{T_\mr{MF}}$ (we set $k_B=1$).

In order to compute expectation values, we define two functions $p\left(\alpha_\mb{r},\alpha_\mb{k}\right)$ and $f\left(\alpha_\mb{r},\alpha_\mb{k}\right)$ as 
  \begin{align}\label{eq:finTObsMC_helper}
   \begin{split}
&p\left(\alpha_\mb{r},\alpha_\mb{k}\right)=\frac{1}{Z} \left|     
     \braket{\alpha_\mb{k}}{\alpha_\mb{r}}\right|^2 \e{-\beta_\mr{MF}E_{\alpha_\mb{k}}}, \\
     & f\left(\alpha_\mb{r},\alpha_\mb{k}\right)=\sum_{\gamma_\mb{r}}\bra{\alpha_\mb{r}}\hat{O}\ket{\gamma_\mb{r}} 
     \frac{\braket{\gamma_\mb{r}}{\alpha_\mb{k}}}{\braket{\alpha_\mb{r}}{\alpha_\mb{k}}}.
   \end{split}
  \end{align}
$p\left(\alpha_\mb{r},\alpha_\mb{k}\right)$ can be interpreted as a probability distribution, such that we can use Metropolis sampling \cite{Metropolis1953} over the space of $\left\{ \left(\alpha_\mb{r},\alpha_\mb{k}\right) \right\}$. The Gutzwiller projection is applied during this Metropolis Monte Carlo sampling procedure by restricting the states $\ket{\alpha_\mb{r}}$ to configurations with at most singly occupied sites.
We thus obtain estimates for the expectation values of an operator $\hat{O}$ depending on the three parameters $\phi$, $T_\mr{MF}/{J_\mr{MF}}$ and ${h}/{J_\mr{MF}}$ as 
  \begin{equation}\label{eq:finTObsMC}
     \expval{\hat{O}} = \Tr\left(\hat{\rho} \hat{O}\right) = \sum_{\alpha_\mb{r},\alpha_\mb{k}}p\left(\alpha_\mb{r},\alpha_\mb{k}\right)f\left(\alpha_\mb{r},\alpha_\mb{k}\right).
  \end{equation}

\section{Optimization Method} \label{sec:opti}

Our goal is to determine the finite temperature RVB state(s) that best describes the experimental data. To this end, we optimize the variational parameters of our trial state such that for a set observables \( \hat{O} \) at distances \( d \) the two- or three-point correlation functions $C_O(d)$ match the experimentally measured correlations as closely as possible. 
We define an objective function $\mathcal{L}$ as the equally weighted sum of squared difference between the estimates $C_O^\mathrm{MC}\left( d \right)$ from our trial state and the experimental or reference values $C_O^\mathrm{ref}\left( d \right)$ for all observables \( \hat{O} \) and distances $d$ of interest: 

  \begin{equation}\label{eq:ObjFun}
    \mathcal{L} = \sum_O \sum_d \left[ C_O^\mathrm{ref}\left( d \right) - C_O^\mathrm{MC}\left( d \right) \right]^2.
  \end{equation}
  
For the optimization, we then estimate the gradients of $\mathcal{L}$ with respect to all variational parameters $\xi_l$ via Metropolis sampling as 

  \begin{equation}\label{eq:GradObj}
    \frac{\partial \mathcal{L}}{\partial \xi_l} = 2 \sum_O \sum_d \left[ C_O^\mathrm{MC}\left( d \right) - C_O^\mathrm{ref}\left( d \right) \right]\frac{\partial}{\partial \xi_l}C_O^\mathrm{MC}\left( d \right).
  \end{equation}
To this end, we evaluate the gradient of the correlation function in Eq.~\eqref{eq:GradObs} below.
The dependence on the parameters shows up either only in the Boltzmann factor in the case of the temperature $T_\mr{MF}$ or in both the Boltzmann factor and the states $\ket{\alpha_\mb{k}}$ in the case of the other two parameters.
We reformulate the derivative of the objective function with respect to the parameter \(\xi_l\) in a similar form as we did in Eq.~\eqref{eq:finTObsMC} and obtain an estimate for the gradient within our MC sampling scheme akin to the variational case \cite{Becca2017}.

  \begin{align}\label{eq:GradObs}
    \begin{split} 
      \frac{\partial}{\partial \xi_l}C_O^\mathrm{MC} &= \lim_{\daq\rightarrow0} \frac{1}{2\daq}\Re{\Tr\sqare{\pare{\dmdaq-\dma}\hat{O}}}\\ 
      & = 2 \Re \left\{ \frac{1}{Z} \sum_{\alpha_{\mb{k}},\alpha_{\mb{r}}}\abs{ \bra{\alpha_{\mb{k}}}\ket{\alpha_{\mb{r}}}}^2 \e{-\beta E_{\mb{k}}} \right.\\
      & \qquad\qquad \left. \sqare{\frac{\bra{\alpha_{\mb{r}}}\hat{O}\ket{\alpha_{\mb{k}}}}{\braket{\alpha_{\mb{r}}}{\alpha_{\mb{k}}}} F\pare{\alpha_{\mb{r}},\alpha_{\mb{k}}}}\right\} \\
      & \approx 2\Re{\frac{1}{N} \sum_i \frac{\bra{\alpha_{\mb{r}}^{(i)}}\hat{O}\ket{\alpha_{\mb{k}}^{(i)}}}{\braket{\alpha_{\mb{r}}^{(i)}}{\alpha_{\mb{k}}^{(i)}}} F\pare{\alpha_{\mb{r}}^{(i)},\alpha_{\mb{k}}^{(i)}}}, \\ 
    \end{split}
  \end{align}
where the sum over $i$ runs over Monte Carlo samples $(\alpha_{\mb{r}}^{(i)},\alpha_{\mb{k}}^{(i)})$ and we define
\begin{align}\label{eq:GradLoc}
\begin{split}
       F\pare{\alpha_{\mb{r}},\alpha_{\mb{k}}} &= \Theta_l\pare{\alpha_{\mb{r}},\alpha_{\mb{k}}}-\Tbar+\frac{1}{2}\pare{\Phi_l\pare{\alpha_{\mb{k}}}-\Pbar} \\
      \text{with } 
    \Theta_l\pare{\alpha_{\mb{r}},\alpha_{\mb{k}}} &= \frac{\partial}{\partial\xi_l} \ln{\braket{\alpha_\mb{r}}{\alpha_\mb{k}}},\\
    \Phi_l\pare{\alpha_{\mb{k}}} &= - \frac{\partial}{\partial\xi_l} \left(\beta E_{\alpha_\mb{k}}\right).
       \end{split}
\end{align}
We define the expectation values as averages over the Monte Carlo samples,

\begin{align}
\begin{split}
    \Tbar &= \left< \Theta_l\pare{\alpha_{\mb{r}},\alpha_{\mb{k}}} \right> \approx \frac{1}{N} \sum_i \Theta_l\pare{\alpha_{\mb{r}}^{(i)},\alpha_{\mb{k}}^{(i)}},\\
    \Pbar &= \left< \Phi_l\pare{\alpha_{\mb{k}}}\right> \approx \frac{1}{N} \sum_i \Phi_l\pare{\alpha_{\mb{k}}^{(i)}}.
    \end{split}
\end{align}

The derivatives in Eq.~\eqref{eq:GradLoc} can be approximated by finite differences, allowing us to obtain estimates of the gradient of the objective function for each parameters. In order to optimize the variational parameters, we then perform gradient descent.

\section{Benchmarks}\label{sec:bench} 

In order to explore the capabilities of our method and understand possible subtleties, we start by looking at several well understood cases.
First we want to identify the ability to recover Monte Carlo parameters of data generated within our Monte Carlo approach.
For this we look at three different cases

\begin{enumerate}
    \item zero temperature square lattice\label{enum:1}
    \item finite temperature square lattice
    \item finite temperature triangular lattice.\label{enum:3}
\end{enumerate}

For case \ref{enum:1} we set \( \tmf=0 \) and for case \ref{enum:3} \(\hmf=0\), while optimizing with respect to the remaining parameters.
We construct the objective function using the spin-spin correlation function averaged over $\mathcal{N}_n$ sites at distance $d$
  \begin{equation}\label{eq:SpSp}
    C_\mathrm{S}\left(\abs{\mb{d}}\right) =
    \frac{1}{\mathcal{N}_\mathrm{n}}\sum_{\mb{i},\mb{d}}\expval{\hat{\mb{S}}_\mb{i}\cdot \hat{\mb{S}}_{\mb{i}+\mb{d}}}.
  \end{equation}
In order to account for short and longer range effects, we consider distances $d=1$ and $d=\sqrt{8}$ for the square lattice and $d=1$ and $d=2$ for the triangular lattice.

\begin{figure}[htbp]
    \centering
    \includegraphics[width=\linewidth]{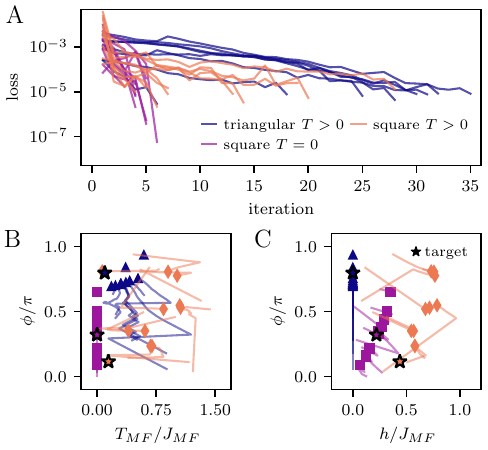}
    \caption{\textbf{Benchmark optimization.} 
             We optimize the parameters of the trial state with respect to expectation values estimated using the same trial state in a zero temperature square lattice (purple squares), a finite temperature square lattice (orange diamonds) and a finite temperature triangular lattice setting (blue triangles).
             In \textbf{A}, we show the convergence of the minimization for the three different parameter sets within a maximum of $\approx 35$ steps up to an order of $10^{-5}$ for the objective function. 
             In panel \textbf{B} and \textbf{C}, we present the evolution of the trial state parameters $\phi$, $\tmf$ and $\hmf$ during the optimization, where each symbol corresponds to the final optimized parameters and the lines to the gradient descent trajectory.
             As expected from Fig~\ref{fig:intro}, we see that it is not possible to recover the exact initial parameters (colored stars).
             Nevertheless, we identify regions in parameter space with expectation values similar to the ones from the initial parameters.%
             }
    \label{fig:rec}
  \end{figure}

In Fig.~\ref{fig:rec} we show that with our method we are capable of minimizing the objective function for this benchmark problem within about 35 optimization steps, depending on the initial parameters, reaching values of orders of at most $10^{-5}$.
For comparison, the nearest-neighbor spin-spin correlation for such systems is of the order of $10^{-1}$.
Comparing the target parameters with the parameters resulting from the optimization we see that the optimized parameters are significantly spread out.
Nevertheless, we are able to identify a region in the parameter space for which the objective function $\mathcal{L}$ is comparably small and, hence, the correlations considered in the objective function are close to the ones of the reference data.
Looking back at Fig.~\ref{fig:intro}A and at ground state calculations in \cite{DallaPiazza2014}, a similar behavior of the objective function or the energy in the parameter space can be observed for this type of trial state, i.e. different sets of variational parameters $\bf{\xi}$ yield very similar values for the correlation functions and the energy.

The second type of reference data we consider stem from quantum Monte Carlo (QMC) simulations \cite{Sandvik1999} of the Heisenberg model on a square lattice, i.e. Hamiltonian \eqref{eq:tJHam} without doping. 
We optimize the SFN state with respect to the RVB parameters \(\tmf,\phi,\hmf\) for different values of $T/J$ in the original QMC simulation ranging from $0.1$ to $1.3$.
Similar to before, we consider spin-spin correlations as defined in Eq. \eqref{eq:SpSp} for distances $d=1$ and $d=\sqrt{8}$ to account for both short- and long-range behavior.
In this setting we optimize all three parameters $\tmf$, $\phi$ and $\hmf$.
We expect that the qualitative behavior of $\tmf$ follows that of $T/J$, meaning that for larger $T/J$ we expect larger values of $\tmf$.
From the objective function landscape shown in Fig.~\ref{fig:intro}A for the half-filled Hubbard model, we expect to see a similar arc-shaped structure depending on $\phi$ and $\tmf$.
In the Heisenberg model the expectation value of the staggered magnetization as well as the range of antiferromagnetic spin-spin correlation decreases with increasing temperature. Hence we expect the parameter $\hmf$, controlling the spin ordering in the system, to decrease with $T/J$.

\begin{figure}[htbp]
    \centering
    \includegraphics[width=\linewidth]{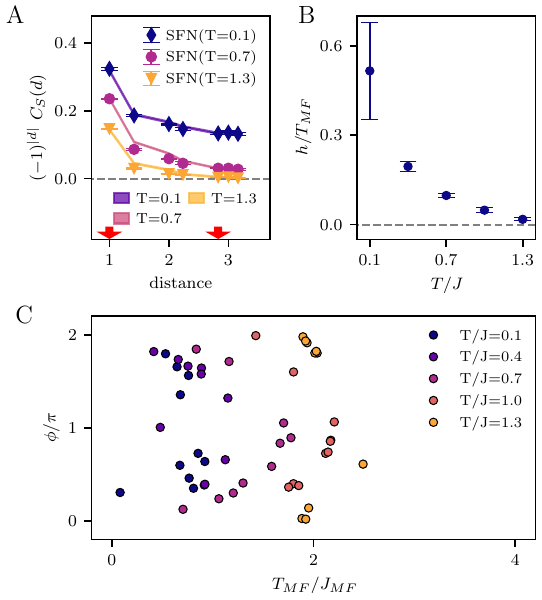}
    \caption{\textbf{Optimization using expectation values from Quantum Monte Carlo calculations.} 
             We present the results of the optimization with respect to expectation values at distances $d=1,~\sqrt{8}$ (red arrows in \textbf{A}) obtained from snapshots from Quantum Monte Carlo simulation \cite{Sandvik1999} of the Heisenberg model at different values for $T/J$.
             In panel \textbf{A}, we show that the optimized states reproduce $C_S$, sign-adjusted for antiferromagnetic correlations, of the QMC simulations (solid lines) at temperatures $T/J=0.1,~0.7,~1.3$.
             We present the ratio $h/T_\mr{MF}$ of the optimized states as a function of $T/J$ in \textbf{B} and show that, as \( h/T_\mr{MF} \) decreases, the Néel field parameter becomes less relevant with increasing \( T/J \).
             \textbf{C} displays the parameters $\phi/\pi$ and $T_{MF}/J_{MF}$ of the optimized states showing an increasing behavior of $\tmf$ with respect to $T/J$.
             }%
    \label{fig:QMC}
  \end{figure}

In Fig.~\ref{fig:QMC} we show the performance in optimizing the correlation functions as well as the behavior of the RVB parameters for the different values of $T/J$.
The optimized trial state is able to capture the behavior of the spin correlation $C_S\left(d\right)$ for all considered values of $d$. Note that we only use $d=1$ and $d=\sqrt{8}$ in the optimization. 
For high temperatures we observe some deviations, probably because of the nearest neighbor correlator dominating the objective function.  

As shown in Fig.~\ref{fig:QMC}B, the ratio of the staggered field with the mean field temperature, $h/T_\mathrm{MF}$, which is the relevant quantity determining the antiferromagnetic ordering in the trial state, exhibits a visible decrease for increasing $T/J$. 

The relation of the parameters $\phi$ and $\tmf$ is similar to previous observations.
At equal values of $\phi$, the state with a higher value of $T/J$ also results in larger values for $\tmf$.
We also observe a similar arc-like structure as seen in the landscape of the objective function in Fig.~\ref{fig:intro}A.
We find some additional peculiarities in the distribution of the parameters:
For small values of $T/J$, the values of the optimized $\phi$ tend to be close to the value of $\phi \approx 0.35\pi$ observed in ground state calculations using a similar trial state with variational Monte Carlo.
For the largest value considered in this paper, $T/J=1.3$, most of the $\phi$ are close to $\phi = 0$ or $\phi = 2 \pi$.
This could be related to previous calculations proposing a state with zero flux for higher temperatures (uniform RVB) \cite{Wen1996}.
Yet, the values of the parameters could to some extent be influenced by the initial parameters of the optimization, which were sampled uniformly from an appropriate region in parameter space.

\section{Square Lattice}\label{sec:square}

Our approach is designed to be used to analyze quantum states realized in experimental setups. After having performed the above benchmark checks on theory data, we can confidently apply the optimization procedure to experimental data from cold-atom simulations of the Fermi-Hubbard model.
For the first dataset \cite{Chiu2019} we analyze the model on a square lattice for different values of hole doping, which we define as $\delta < 0$, using both spin-spin correlators
\begin{equation}
    C_\mathrm{Z}\left(\abs{\mb{d}}\right) =
    \frac{1}{\mathcal{N}_\mathrm{n}}\sum_{\mb{i},\mb{d}}\expval{\hat{S}_\mb{i}^{z}\hat{S}_{\mb{i}+\mb{d}}^{z}}_\mathrm{C}
    \label{eq:SzSz}
\end{equation}
and dopant-dopant correlators 
  \begin{equation}
    C_\mathrm{D}\left(\abs{\mb{d}}\right)=
    \frac{1}{\mathcal{N}_\mathrm{n}}\sum_{\mb{i},\mb{d}}\expval{\hat{n}_{D,\mb{i}}\hat{n}_{D,\mb{i}+\mb{d}}}_\mathrm{C},
    \label{eq:DD}
  \end{equation}
where $\hat{n}_{D,\mb{i}}$ is the dopant number operator at site $\mb{i}$. We consider reference values for the correlators at distances $d=1$, $\sqrt{2}$ and $2$ for $C_Z$ and $d=\sqrt{2}$ for $C_D$ from the experimental data.
In the Fermi-Hubbard model, doublon-hole fluctuations exist, which are not captured by our trial state due to the Gutzwiller projection. In the imaging procedure in the experiments considered here, such virtual doublon hole pairs are mapped to two empty sites. At the values of $U/t$ considered here, doublon hole pairs almost exclusively occur on nearest neighboring sites, i.e. $d=1$, and we thus do not consider this distance for $C_D$. 

The parameters of the model in the experiment \cite{Chiu2019} are $U/t = 8.1$ with an estimated temperature of $T/J \approx 0.65$.
The second dataset, from Ref.~\cite{Xu2023}, includes spin-spin correlation functions $C_Z$ for distances $d=1$ and $d=\sqrt{2}$ for dopings ranging from about $-0.5$ to $0.55$, i.e. for hole as well as particle doping. 
The model parameters in this case are $U/t \approx 9$ at a temperature of $T/J \lessapprox 0.8$. We perform a separate optimization procedure for each dataset and within each dataset for each doping value, thus allowing a doping dependence of the variational parameters of our optimized RVB state.

For the experimental data, we optimize the SF state with respect to the parameters $\phi$ and $\tmf$ for the correlation functions mentioned above for a maximal number of 300 optimization steps. At the temperatures $T/J \approx 0.65...0.8$ considered here, we have seen in Sec.~\ref{sec:bench} that the spin-spin correlations are sufficiently short-ranged to yield a very small value of the staggered magnetic field $h/T_\mr{MF}$ in the optimization procedure, and we thus set $\hmf=0$ here. 

  \begin{figure*}[htbp]
    \centering
    \includegraphics[width=\linewidth]{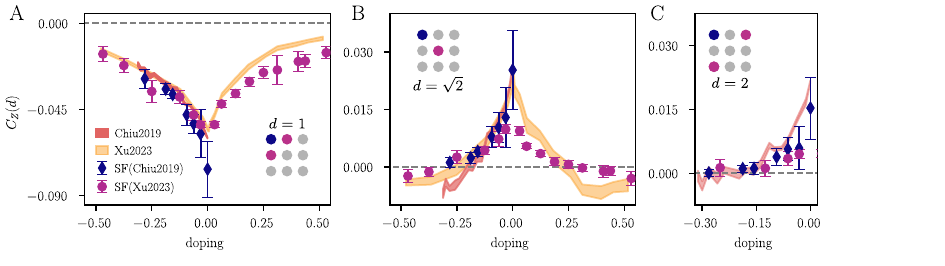}
    \caption{\textbf{Optimized correlations $C_Z$ for the square lattice Fermi-Hubbard model.} 
              We minimize the objective function with respect to $C_Z\left(d\right)$ 
              using correlations from \cite{Chiu2019} and \cite{Xu2023}
              at distances $d=1$, $d=\sqrt{2}$ and for \cite{Chiu2019} also $d = 2$.
             We observe that the SF state captures \(C_Z(d)\) well for all distances considered, even if a given distance is not included in the optimization as for \(d=2\) and \cite{Xu2023}.
              Our ansatz cannot accurately capture the sign change for $d=\sqrt{2}$ for large dopings, especially for the data from \cite{Chiu2019}.
              Some of the larger errors, e.g. at half-filling might be attributed to non-converged optimizations.}
    \label{fig:sqSzSz}
  \end{figure*}

In Fig.~\ref{fig:sqSzSz} we show the correlations $C_Z$ for the optimized states averaged over all runs in comparison to the reference values. Per doping we performed ten runs with different initial parameters for each doping value and the errorbar shows the standard deviation of \(C_Z\) of the runs.
Overall, our ansatz is capable of capturing the behavior of the correlations from the experiments.
Yet, for some dopings the error is quite large. 
For example at half-filling, this can be attributed to non-converged optimization runs due to the structure of the optimization landscape shown in Fig.~\ref{fig:intro}A.
The most significant difference in \(C_Z\) between the RVB state and the experiment is the sign change for $d=\sqrt{2}$, Fig.~\ref{fig:sqSzSz}B, which is not captured correctly by our method.
The decreasing performance of our method for longer-range interactions can to some extent be attributed to the dominance of stronger correlations in our objective function, which in this case favors nearest neighbor correlations, and the effect of doublon-hole pairs, which reduce the correlator measured for $d=1$.

\begin{figure}[htbp]
    \centering
    \includegraphics[width=\linewidth]{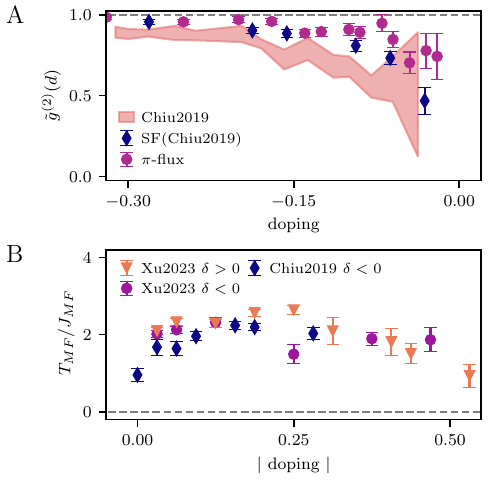}
    \caption{\textbf{Hole-hole correlations and behaviour of $T_{MF}/J_{MF}$ with doping.} 
             In \textbf{A}, we see the function $\Tilde{g}^{\left(2\right)}$, measuring the hole-hole correlation for the square lattice Fermi-Hubbard model at a distance of $d=\sqrt{2}$. The numerical results from the optimization qualitatively agree with the experimental data and follows the doping behavior better than the \(\pi\)-flux state considered in \cite{Chiu2019}.
             Panel \textbf{B} depicts the relation between the mean value of the parameter $\tmf$ and the doping, showing, as expected for the square lattice, a similar behaviour for particle and hole doping. We see that the mean-field temperature slightly increases with the doping up to around $0.25$ after which a decline can be observed upon further increase.
             }%
    \label{fig:sqDop}
  \end{figure}

The optimized states also capture the dopant correlation $C_D\left(d=\sqrt{2}\right)$ well for the dataset from \cite{Chiu2019}.
As shown in Fig.~\ref{fig:sqDop}A, we see that for the function $\tilde{g}^{(2)}$, defined as 
  \begin{equation}\label{eq:Holeg2}  \tilde{g}^{\left(\mathrm{2}\right)}\left(\abs{\mb{d}}\right)=
    \frac{1}{\delta^2}C_\mathrm{D}\left(\abs{\mb{d}}\right)+1,
  \end{equation}
the RVB state agrees well with the experimental observations.
Comparing this to the non-optimized $\pi$-flux states, where $\phi = \pi$ and $\tmf = T/J$ have been fixed, as considered in \cite{Chiu2019}, our ansatz performs significantly better.

In Fig.~\ref{fig:sqDop}B, we show the doping dependence of the effective temperature $\tmf $. On the square lattice, doublon and hole doping are equivalent and we thus consider the dependence on the absolute value of doping $\delta$. Intuitively, the Fermi-Hubbard model at high dopings $|\delta| \approx 50 \%$ resembles free fermions due to the low overall filling. In terms of the mean field Hamiltonian for our ansatz, Eq.~\eqref{eq:MFHam}, the physics of free fermions is captured when we set $J_\mr{MF} = t$, the original hole hopping. As can be seen in the figure, the mean field temperature in units of $J_\mr{MF}$ decreases from $\approx 30\%$ doping onwards, as expected if the mean field coupling $J_\mr{MF}$ effectively increases towards $t$. At half-filling, on the other hand, the Hubbard model is well described by the Heisenberg model, which is captured by our RVB ansatz for $J_\mr{MF}  = J = 4t^2/U$. The optimized mean field temperature is thus higher at small dopings and decreases in the large doping regime. 

\section{Triangular Lattice}\label{sec:tri}

  \begin{figure}[htbp]
    \centering
    \includegraphics[width=\linewidth]{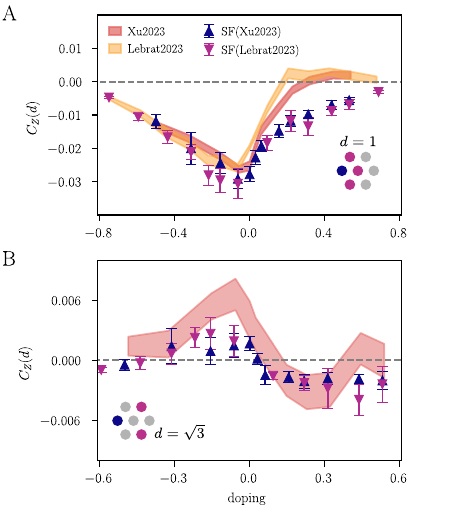}
    \caption{\textbf{$C_Z$ in the triangular lattice Fermi-Hubbard model.} 
             We optimize our ansatz with respect to \cite{Lebrat2024} and \cite{Xu2023}.
             For the triangular lattice, the spin-spin correlation defined in Eq.~\eqref{eq:SzSz} shows a particle-hole asymmetry. Within our ansatz, we are able to capture some part of this asymmetry, like the overall shape as well as the different signs for $d=\sqrt{3}$ for low doping. We are unable to capture the behavior of the spin correlations accurately for large particle doping. The sign change above a particle doping of around $0.3$ for $d=1$ and the behavior for $d=\sqrt{3}$ around the same doping are not observed with our RVB ansatz. For hole doping we are able to find fitting parameters to describe the experimentally observed correlations very well across the entire doping range. Note that for $d=\sqrt{3}$, there is no experimental data available for the highest shown hole doping. The values at these dopings correspond to the optimization w.r.t only the $d=1$ spin correlations and \(C_{DZZ}\). 
             }
    \label{fig:trSzSz}
  \end{figure}

Anderson's original idea was that RVB states capture the physics of the Heisenberg model on the geometrically frustrated triangular lattice \cite{Anderson1973}. While the ground state with just nearest neighbor spin exchange has been shown to be a 120 degree ordered state, additional couplings like next-nearest-neighbor spin exchange \cite{Drescher2022}, or ring-exchange terms, can potentially lead to a quantum spin liquid ground state \cite{Motrunich2005,Gong2017,Cookmeyer2021,Chen2022_QSL}. Such terms can arise in higher order in $U/t$ when going from the Hubbard to $t-J$-type models, and can potentially be induced through the motion of dopants away from half-filling \cite{Schloemer2023_rec}. 
Variational Monte Carlo using RVB type ans\"atze have shown qualitative agreement with numerically determined ground state phase diagrams of Heisenberg-type models with additional or anisotropic couplings, including the presence of quantum spin liquid states \cite{Ghorbani2016,Iqbal2016,Weber2006}.

In the present work, we consider finite temperature states, where no long-range order is observed in the two-dimensional system, and RVB states potentially provide a good description of the local physics. 
    In recent experiments, triangular lattices have been implemented using ultracold atoms \cite{Yang2021,Mongkolkiattichai2023,Xu2023,Lebrat2024,Prichard2023}. The different correlation functions obtained in these experiments can be used within our method to obtain the optimized RVB state describing the actual system.
    We consider two different experimental realizations and use $C_Z$ for $d=1$ and $d=\sqrt{3}$ from Ref.~\cite{Xu2023} and $C_Z$ for $d=1$ and $C_{DZZ}$, defined as

\begin{multline}\label{eq:trDZZdop}
     C_\mathrm{DZZ}\left(\delta\right) =~\frac{1}{3L}\sum_\mb{i} \left[ C_\mathrm{DZZ}\left(\mb{i};\mb{e}_1,\mb{e}_2\right) + \right.\\
     \left.C_\mathrm{DZZ}\left(\mb{i};\mb{e}_2,\mb{e}_3\right)+C_\mathrm{DZZ}\left(\mb{i};\mb{e}_3,-\mb{e}_1\right)\right]
\end{multline}
  with system size $L$ and the connected dopant-spin-spin correlation
    \begin{equation}\label{eq:trDZZdis}
C_\mathrm{DZZ}\left(\mb{i};\mb{d}_1,\mb{d}_2\right)=\frac{1}{\mathcal{N}_\mathrm{dss}} \expval{\hat{n}_{D,\mb{i}}\hat{S}_{\mb{i}+\mb{d}_1}^{z}\hat{S}_{\mb{i}+\mb{d}_2}^{z}}_\mathrm{C},
  \end{equation}
 from Ref.~\cite{Lebrat2024} as reference data.
 The number operator for dopants is $\hat{n}_{D,\mb{i}}$ and the normalization $\mathcal{N}_\mr{dss}$ is the probability to find a dopant at site $\mb{i}$ and two spins at sites $\mb{i}+\mb{d_1}$ and $\mb{i}+\mb{d_2}$.
 The vectors \(\mb{e}_j\) denote the neighboring sites of \(\mb{i}\).
 In Ref.~\cite{Xu2023}, $U/t = 9.2$ and $T/t =0.39$, whereas the data used from Ref.~\cite{Lebrat2024} was obtained at $U/t = 20.6$ and $T/t = 0.30$. 
 The SF state is optimized with respect to the parameters \(\tmf\) and \(\phi\). 

    In Fig.~\ref{fig:trSzSz} we plot $C_Z$ as a function of the doping for distances $d=1$ and $d=\sqrt{3}$.
    The data from both experiments show a similar doping dependence, including a clear asymmetry between particle and hole doping for both distances as well as a sign change for large particle doping.
    Looking at the values of $C_Z$ for the optimized states, we see a similar doping behavior regardless of the correlation considered in the objective function. 
    In both cases the optimized values show better agreement with the experimental data for hole doping than for particle doping.
    In the latter case, our ansatz does not capture the slightly ferromagnetic correlations correctly, as we still observe antiferromagnetic correlations for all considered dopings at distance $d=1$.
    Due to the small values for $d=\sqrt{3}$, it is difficult to make more substantial claims, yet the optimized data qualitatively agrees with the experiments.

\begin{figure}[htbp]
    \centering
    \includegraphics[width=\linewidth]{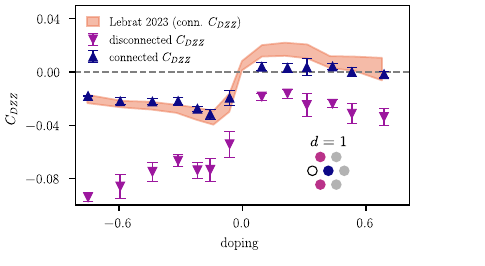} 
    \caption{\textbf{Spin-correlation around dopant $C_{DZZ}$ in the triangular lattice Fermi-Hubbard model.}
             Optimizing with respect to both the spin-spin correlations and measurements of $C_{DZZ}$, we observe that the spin-spin correlation next to a dopant $C_{DZZ}$ as defined in Eq.~\eqref{eq:trDZZdop} can be qualitatively captured by our ansatz. While we see good performance for hole doping, we are not able to find parameters for our RVB ansatz to accurately match the experimentally measured $C_{DZZ}$ in the intermediate particle doped regime \cite{Lebrat2024}.
             It is important to note that the ferromagnetic correlations for the connected \(C_{DZZ}\) are not present for the disconnected correlator.
             }
    \label{fig:trDZZ}
  \end{figure}

    In Fig.~\ref{fig:trDZZ} we show the connected dopant-spin-spin correlation function $C_{DZZ}$ as well as its disconnected version. This correlator describes the spin correlation neighboring a dopant, which has been analyzed to capture the local physics of magnetic polarons in the case of the square lattice  \cite{Koepsell2019,Koepsell2021} and the triangular lattice \cite{Prichard2023,Lebrat2024}. 
    For the connected case, the experimental measurements show antiferromagnetic correlations for hole doping and ferromagnetic correlations for particle doping, resulting in a sign change at half-filling.
    This doping behavior can be reproduced qualitatively using our ansatz by optimizing w.r.t. $C_Z\pare{d=1}$ and $C_{DZZ}$. While the hole doped regime is quantitatively well described by our ansatz, there are some minor deviations between the RVB state and the experiment on the particle doped side.
    In constrast to that the disconnected correlator shows no signs of fermionic correlation. Thus, the fermionic correlation seen for \(C_{DZZ}\) is a consequence of subtracting the spin-spin correlation from the disconnected part.

    Our analysis shows that for hole doping, the behavior of all considered correlation functions can be captured by our ansatz, whereas the sign change in $C_Z\pare{d=1}$ as well as details of the behavior of $C_{DZZ}$ for particle doping cannot be described by the RVB ansatz.

\section{Summary and Outlook}\label{sec:sum}

In this paper we have used a gradient descent based optimization procedure to determine the variational parameters of a finite temperature RVB ansatz state that best describes experimental data of cold atom realizations of the Fermi-Hubbard model in the square and triangular lattice. Despite the simplicity of our class of ansatz states, we observer that it is capable of capturing the behavior in spin-spin, dopant-dopant, and third order dopant-spin-spin correlations of the experimental data. Notably, for the experimental data, we only use two variational parameters -- the staggered flux $\phi$ and the effective temperature $\tmf$ -- compared to hundreds of thousands of parameters in typical numerically optimized variational states such as matrix product states or neural quantum states. Upon optimizing the variational parameters through comparison with the experimental data, we are able to qualitatively capture most of the non-trivial behavior of the considered correlation functions, like the asymmetry between particle and hole doping in the triangular lattice. Compared to the non-optimized $\pi$-flux state, based on the same ansatz considered here, but with fixed instead of optimized variational parameters, used in Ref.~\cite{Chiu2019}, we obtain significantly better agreement in the charge sector and fully capture the doping dependence of the $\Tilde{g}^{(2)}(d=\sqrt{2})$ function. Apart from this better agreement, another advantage of the optimization of the variational parameters introduced in this work is the resulting doping dependence of the variational parameters, which can be directly interpreted to gain physical insights.

An interesting direction for future work consists in improving the ansatz as well as the optimization procedure. One possibility would be to include (tightly bound) doublon-hole pairs in the ansatz \cite{Gutzwiller1963}, as the appearance of doublon-hole pairs in the experimental data and the lack thereof in our trial state has led to discrepancies as seen for the spin correlations.
In order to make best use of the available data in the form of quantum projective measurements, or snapshots, it is desirable to include information beyond the specific second- and third-order correlations used in the present work in the optimization procedure, e.g. by minimizing the Kullback-Leibler divergence between the experimental dataset and snapshots sampled from the RVB state. 
The optimal RVB parameters, obtained in this work through comparison of RVB states with experimental data, can furthermore be used as initial states e.g. for ground state searches in hybrid experimental-numerical schemes \cite{Bennewitz2022,Czischek2022,Moss2024,Lange2024hybrid}. 

Apart from the square and triangular lattice Fermi-Hubbard model considered here, other lattice geometries and models, for instance spin-1/2 Heisenberg or transverse field Ising type models, can be studied using the same type of Ansatz and optimization procedure. To this end, quantum simulation data from different platforms, e.g. Rydberg atoms in tweezer arrays, where quantum spin liquid-like states have been realized \cite{Semeghini2021}, can be used. 
Additionally, the analysis of $t-J$ and Fermi-Hubbard type models has shown that after properly integrating out the charge degrees of freedom, the spin sector of these systems can show quantum spin liquid-like behavior \cite{Schloemer2023_rec}. The corresponding data can be directly analyzed with the methods proposed here, enabling the analysis of a possible hidden resonating valence bond state scenario, akin to the original idea that the hole motion induced frustration in the spin sector leads to an RVB state at finite doping \cite{Edegger2007}. 

Finally, different classes of systems can potentially be analyzed using the technique introduced here in combination with other types of RVB states, such as those relevant to van der Waals materials \cite{Koenig2022}.  
\\

\textbf{Acknowledgements.--}
We thank Aaron Young and Hannah Lange for fruitful discussions. This research was funded by the Deutsche Forschungsgemeinschaft (DFG, German Research Foundation) under Germany's Excellence Strategy -- EXC-2111 -- 390814868, by the European Research Council (ERC) under the European Union’s Horizon 2020 research and innovation programm (Grant Agreement no 948141) — ERC Starting Grant SimUcQuam, by the NSF through a grant for the Institute for Theoretical Atomic, Molecular, and Optical Physics at Harvard University and the Smithsonian Astrophysical Observatory, by the AWS Generation Q Fund at the Harvard Quantum Initiative (YG) and in part by the Austrian Science Fund (FWF) [10.55776/COE1]. LHK acknowledges funding from the NSF Graduate Research Fellowship Program. AK acknowledges funding from the NSF Graduate Research Fellowship Program. CR acknowledges funding from Cusanuswerk, Bischöfliche Studienstiftung, Germany. For Open Access purposes, the author has applied a CC BY public copyright license to any author accepted manuscript version arising from this submission.

\textbf{Data Availabilty.--}
The data that support the findings of this article are openly available \cite{reinmoser_2025_16367725}.

\bibliography{main}

\end{document}